\documentclass[a4paper,10pt]{article}
\usepackage{
times,
amsmath,amssymb,epsfig,graphics,graphicx}

\newcommand{\e}{\textrm { e}}

\newcommand{\be}{\begin{eqnarray}}
\newcommand{\ee}{\end{eqnarray}}

\begin{document}

\begin{center}
{\Large\bf  Angular Momentum Dependent Quark Potential of QCD Traits
and Dynamical $O(4)$ Symmetry  }
\end{center}
\vspace{0.5cm}

\begin{center}
{\Large C.\ B.\ Compean and M.\ Kirchbach\footnote{Invited talk at
the Mini-Workshop Bled 2006, 
 ``Progress in Quark Models'', Bled, Slovenia, July 10-17, 2006 }}
\end{center}

\vspace{0.5cm}
\begin{center}
{\it Instituto de F\'{\i}sica}, \\
         {\it Universidad Aut\'onoma de San Luis Potos\'{\i},}\\
         {\it Av. Manuel Nava 6, San Luis Potos\'{\i}, S.L.P. 78290, M\'exico}
\end{center}

\vspace{0.5cm}

\begin{flushleft}
{\bf Abstract:}
A common quark potential that captures the essential traits of
the QCD quark-gluon dynamics is expected to 
(i) interpolate between a Coulomb-like
potential (associated with one-gluon exchange) and the infinite wall
potential (associated with trapped but  asymptotically free quarks), 
(ii) reproduce in the intermediary region  the linear
confinement  potential (associated with multi-gluon self-interactions)
as established by lattice QCD calculations of hadron properties.
We first show that the exactly soluble trigonometric Rosen-Morse 
potential possesses all these properties. 
Next we observe that this potential, once interpreted as 
angular momentum dependent, acquires a dynamical $O(4)$ symmetry and 
reproduces exactly  quantum numbers and level splittings of the non-strange
baryon spectra in the $SU(2)_I\otimes O(4)$  classification 
scheme according to which baryons cling on to multi-spin parity clusters
of the type $\left(\frac{K}{2},\frac{K}{2} \right)\otimes
\lbrack \left( \frac{1}{2},0\right) \oplus \left( 0, \frac{1}{2}\right)
\rbrack$, whose relativistic image is $\psi_{\mu_{1}...\mu_{K}}$.
Finally, we bring exact energies and wave functions of the levels
within the above potential and thus put it on equal algebraic footing
with such common potentials of wide spread as are
the harmonic-oscillator-- and the Coulomb potentials.  

\end{flushleft}

\vspace{0.3cm}
\begin{flushleft}
\end{flushleft}

\section{Introduction}
The non-strange baryon spectra below $\sim 2500$ MeV
reveal, isospin by isospin, as a striking  phenomenon  
mass degenerate series of $K$ pairs of resonances of opposite
spatial parities and spins ranging from
$\frac{1}{2}^{\pm}$ to $\left( K- \frac{1}{2}\right)^{\pm}$ 
which terminate by a highest spin--$\left( K+\frac{1}{2}\right)$ 
resonance that remains unpaired \cite{MK-97}. Such series
(displayed in Fig.~1) perfectly fit into $SU(2)_I\otimes O(4)$ 
representations of 
the type $\left(\frac{K}{2},\frac{K}{2} \right)\otimes
\lbrack \left( \frac{1}{2},0\right) \oplus \left( 0, \frac{1}{2}\right)
\rbrack$, an observation due again to \cite{MK-97}.
\begin{figure}
{(a)}\includegraphics[width=70mm,height=70mm]{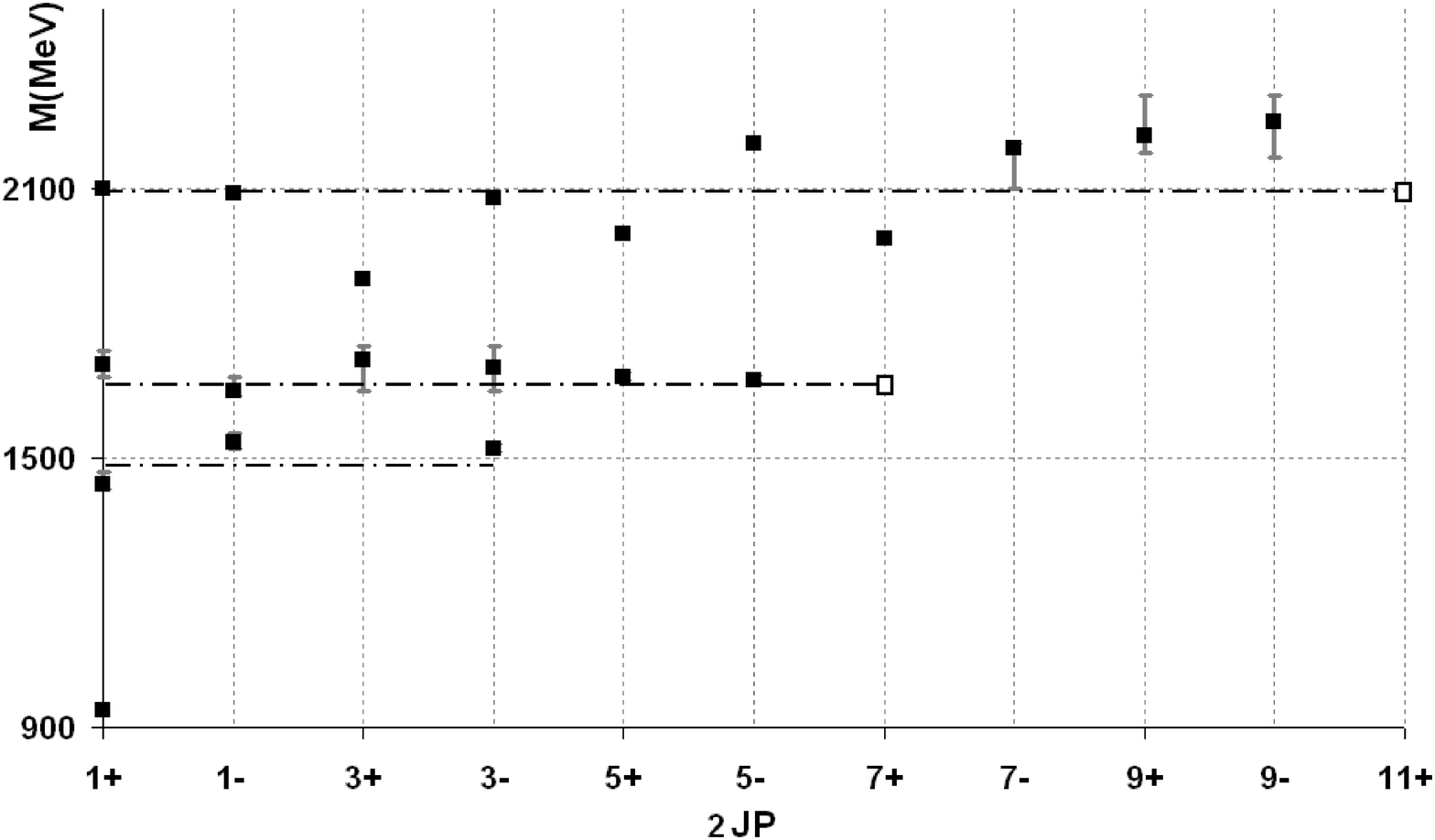}
{(b)}\includegraphics[width=70mm,height=70mm]{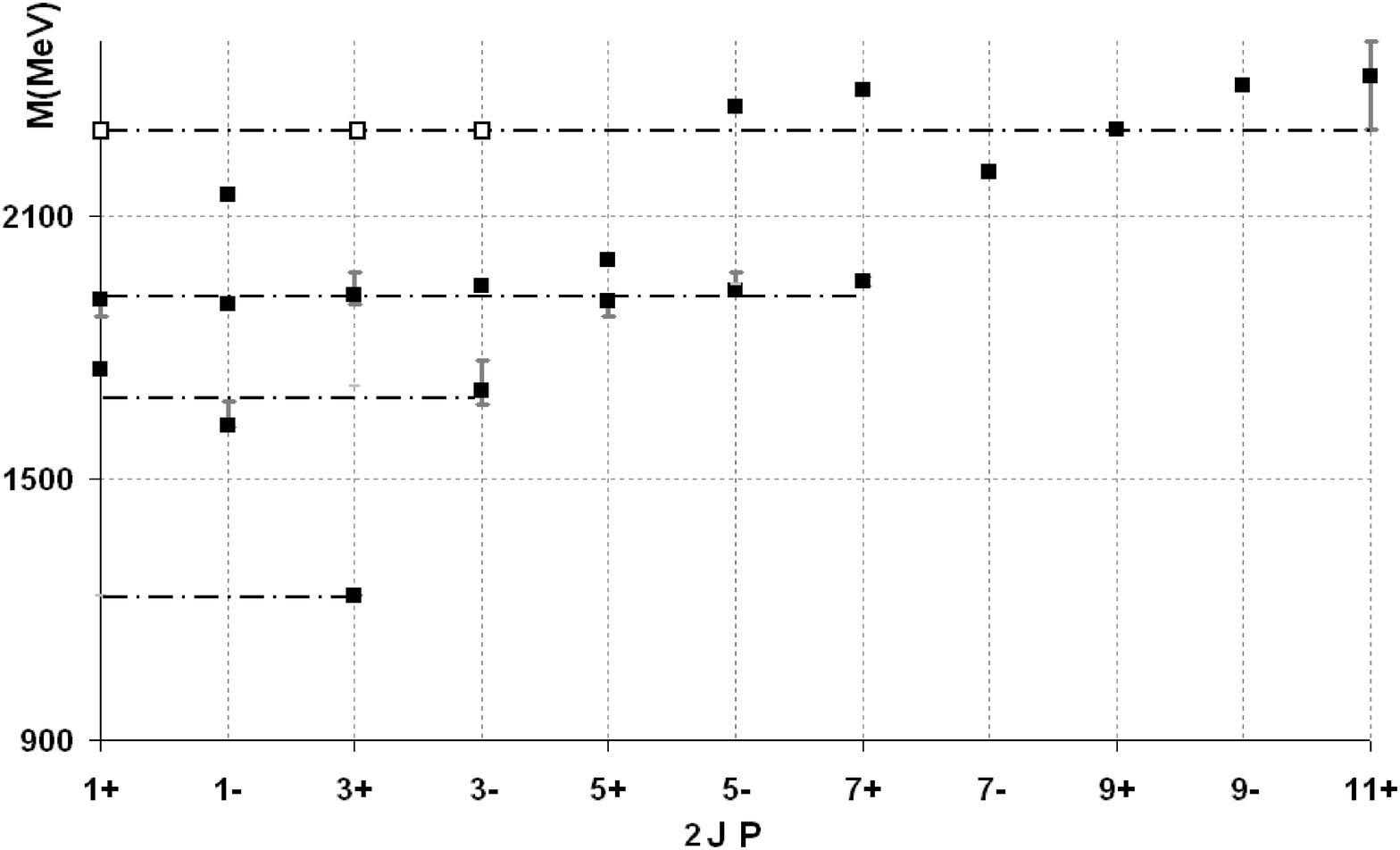}
\caption{Experimentally observed baryon resonances (l.h.s.) $N$ and (r.h.s.) 
$\Delta$. The dash-point lines represent the series mass average. 
Notice that the resonances
with masses above 2000 MeV are of significantly lower 
confidence but those with
masses below 2000 MeV where the degeneracy is very well pronounced. Empty 
squares denote predicted (``missing'') states. Typical, the {\tt \bf 
systematical lack of a parity partner to the first highest spins } 
$D_{I3}$, $F_{I7}$, and $H_{I, 11}$ (the last two being among the 
``missing'' $N$ states).
}
\end{figure}
The appeal of the above classification is twofold.
On the one side, up to the $\Delta (1600)$ state which is likely to be 
a hybrid, no other resonances drop out of the proposed scheme.
Also the prediction of less ``missing'' resonances relative to others
schemes falls under this issue.
On the other side, due to the local $O(4)\sim O(1,3)$ isomorphism,
the  non-relativistic $O(4)$ multiplets  have as an exact relativistic
image the covariant high-spin degrees of freedom given by the 
totally symmetric rank-$K$ Lorentz tensors with Dirac
spinor components, $\psi_{\mu_{1}...\mu_{K}}$ 
known as spin-$\left( K+\frac{1}{2}\right)$
Rarita-Schwinger fields \cite{RS}.
In this fashion, one can view the series of mass degenerate
resonances of alternating parities and 
spins rising from $\frac{1}{2}^\pm $ to $\left(K+\frac{1}{2}\right)^P $ 
as rest frame $\psi_{\mu_{1}...\mu_{K}}$ of mass $m$ and look for
possibilities to generate such multiplets as bound states within an
appropriate quark potential.
Although the degeneracy of the non-strange baryons follows same
patterns as the states of an electron with spin in the Hydrogen atom, 
the  level splittings are quite different. 
The mass formula that fits the $N(\Delta )$ spectrum 
has been encountered in Ref.~\cite{MK_2000} as
\begin{eqnarray}
M_{(\sigma ;I)}-M_{(1;I)} = -f_{I}\frac{1}{\sigma ^{2}}+ 
g_{I}\frac{\sigma ^{2}-1}{4}\, ,
&\quad& \sigma =K+1, \quad I=N, \Delta\, ,
\label{mass-fla}\\
f_{N}=f_{\Delta}=600\,\, \mbox{MeV}, \quad  g_N =70\,\, \mbox{MeV}, &\quad &
g_\Delta =40\,\, \mbox{MeV}\, ,
\label{mfla_prms}
\end{eqnarray}
and contains besides the Balmer-like term, $\left( \sim -1/\sigma ^2\right)$, 
also its inverse.
In effect, the baryon mass splittings increase with $\sigma $.
The degeneracy patterns and the mass formula have found explanation
in Ref.~\cite{KiMoSmi}  within
a version of the interacting boson model (IBM) for baryons. 
To be specific,  to the extent QCD prescribes baryons to be 
constituted of three quarks in a color singlet state, one can 
exploit for  the description of baryonic systems
algebraic models developed for the purposes of triatomic molecules,
a path pursued by  Refs.~\cite{U(7)}.
An interesting dynamical limit of the three quark system is 
the one where two of the quarks 
are ``squeezed'' to one independent entity, a di-quark (qq), 
while the third quark (q) remains spectator. In this limit,
which corresponds to
$U(7)\longrightarrow U(3)\times U(4)$, one can exploit the following chain 
of reducing $U(4)$ down to $O(2)$  
\begin{eqnarray}
&&U(4)\supset O(4)\supset O(3)\supset O(2)\, ,\nonumber\\
&&N\qquad\quad K\qquad\quad  l\qquad\quad m\,  
\label{chains}\\
&& K=N, N-2, ... 1(0), \quad l=K,K-1, ..., 0, \quad |m|< l\, ,
\label{brnsh_rules}
\end{eqnarray}
in order to describe the \underline{ro}tational and 
\underline{vibr}ational (rovibron) modes of the $(qq)-q$  dumbbell.
In so doing, one reproduces the quantum numbers describing 
the degeneracies in the light quark baryon spectra
by means of the following Hamiltonian:
\begin{eqnarray}
{\mathcal H}-{\mathcal H}_0&=&-f_I(4{\mathcal C}_2(so(4))+1)^{-1}+
g_I{\mathcal C}_2(so(4))\, ,\\
{\mathcal C}_2(so(4))\left( \frac{K}{2},\frac{K}{2}\right)&=&
\frac{(K+1)^2 -1}{4}\left( \frac{K}{2},\frac{K}{2}\right)\, .
\label{O(4)_Ham}
\end{eqnarray}
with ${\mathcal C}_2(so(4))$ being the second $so(4)$ Casimir operator. 
In the second row of Eq.~(\ref{chains}) we indicate the quantum
numbers associated with the respective group of the chain.
Here, $N$ stands for the principle quantum number of the four dimensional
harmonic oscillator associated with $U(4)$,
$K$ refers to the $O(4)$ four dimensional angular momentum,
while $l$, and $m$ are in turn ordinary three-- and two angular momenta.
In Ref.~\cite{KiMoSmi}  the interested reader can find all the details
of the algebraic description of the nucleon and $\Delta$ resonances
within the rovibron limit.

Yet, as a principle challenge still remains finding  
a suitable quark potential 
that leads to the above scenario.
In the present work we make the case that the exactly soluble
trigonometric Rosen-Morse potential is precisely the potential we 
are looking for.

The paper is organized as follows. In the next section we analyze the 
shape of the trigonometric Rosen-Morse potential.
In section III we present the exact real orthogonal polynomial 
solutions of the corresponding Schr\"odinger equation. The paper
ends with a brief concluding section.

\section{The shape of the trigonometric Rosen-Morse potential }

We adopt the following form
of the trigonometric Rosen-Morse potential \cite{Sukumar},\cite{Khare} 
\begin{equation}
 v(z)=-2 b \cot z +a(a+1)\csc^2 z\, , \quad a>-1/2, 
\label{v-RMt}
\end{equation}
displayed in  Fig.~2. Here,  
\begin{eqnarray}
z=\frac{r}{d},\quad v(z)=V(z)/(\hbar^{2}/2md^{2}),
\quad \epsilon_n &=&E_n/ (\hbar^{2}/2md^{2})\,,
\end{eqnarray}
the one-dimensional variable is $r=|\mbox{r}|$, 
$ d$ is a properly chosen length scale, $V(r)$ is the potential in 
ordinary coordinate space, and $E_n$ is the energy of the levels.
Our point here is that $v(z)$ interpolates between a Coulomb-like and
an infinite-wall potential going through an intermediary region of
linear, and quadratic (harmonic-oscillator) dependences in $z$.
To see this (besides inspection of
Fig.~2) it is quite instructive to expand the potential
in a Taylor series which for appropriately small $z$,
takes the form of a Coulomb-like 
potential with a centrifugal-barrier like term (if $a$ 
were to be a positive integer)
provided by the $\csc^2 z$ part,
\begin{eqnarray}
v(z)\approx -\frac{2b}{z} +\frac{a(a+1)}{z^2}\, .
\label{Taylor_Coul}
\end{eqnarray}
In an intermediary range where inverse powers of $z$ may be neglected,
one finds the linear plus a harmonic-oscillator potentials
\begin{equation}
v(z)\approx
\frac{2b}{3}\, z +\frac{a(a+1)}{15}z^2\, .
\label{Taylor_lin_HO}
\end{equation}
Finally, as long as  $\cot\, z \stackrel{z\to \pi}{\longrightarrow}
\infty$ and $\csc^2\, z \stackrel{z\to \pi}{\longrightarrow}\infty$,
the potential obviously evolves to an infinite wall.
\begin{figure}
\begin{center}
\includegraphics[width=70mm,height=70mm]{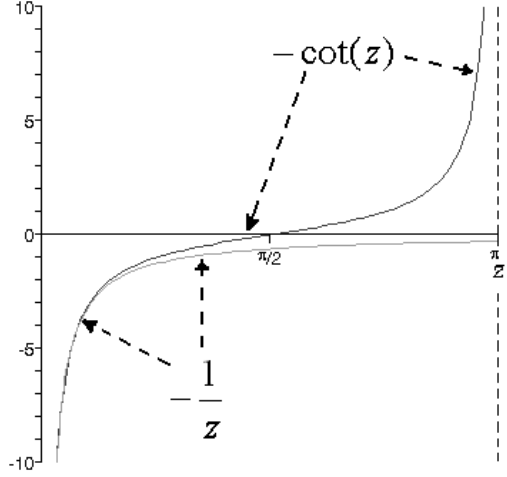}
\caption{
The proximity of the $(\sim \cot z )$- to the $\left( \sim \frac{1}{z}\right) $ 
potential.}
\end{center}
\end{figure}
The above shape captures surprisingly well the essentials of the QCD 
quark-gluon dynamics where the one gluon exchange gives rise to 
an effective Coulomb-like potential, while the self-gluon interactions
produce a linear confinement potential as established  by lattice calculations
of hadron properties.
Finally, the infinite wall piece of the trigonometric
Rosen-Morse potential provides the regime suited for  
trapped but asymptotically free quarks.
By the above considerations one is tempted to conclude that 
the potential under consideration may be a good candidate for a 
common quark potential of QCD traits.

\noindent
In the next section we present the exact solutions of the
Schr\"odinger equation with the trigonometric Rosen-Morse potential.

\section{Energies and wave functions of the levels within the
trgonometric Rosen-Morse potential}
The three-dimensional  Schr\"odinger equation with the trigonometric
Rosen-Morse potential (tRMP) reads:
\begin{equation}
\vec \nabla^{\, 2} \ \psi (\mathbf{z}) +\left(
2 b\cot z-\frac{a(a+1)}{\sin^2 z}+\epsilon\right) \psi ({\mathbf z})=0\, ,
\label{Sch-RMt}
\end{equation}
and is solved in polar coordinates in the standard way by separation
of variables. In effect, the wave function factorizes according to
\begin{equation}
\psi ({\mathbf z})=\frac{R(z)}{z}Y_{l}^{\mu }(\theta,\phi ), 
\quad l=0,1,2,..., \quad
|\mu |< l,
\label{wafu_psi}
\end{equation}
where $Y_l^\mu(\theta ,\phi )$ stand for the standard spherical
harmonics, and $R(z)$  satisfies the one-dimensional equation
\begin{equation}
\frac{d^{\, 2}}{d^2z} \ R (z) +\left(
2 b\cot z-\frac{a(a+1)}{\sin^2 z}+\epsilon\right) R ( z)=0\ .
\label{wafu_U(r)}
\end{equation}
This equation (up to inessential notational differences) 
has been solved in our previous work ~\cite{CK}.
There, we exploited the following factorization ansatz
\begin{equation} 
R (z) =\e^{-\alpha z/2} (1+\cot ^2 z)^{\frac{-(1-\beta )}{2}}
C^{(\beta , \alpha  )} (\cot z)\, , 
\label{Schroed}
\end{equation}
with $\alpha$ and $\beta$ being constant parameters.
Upon introducing the new variable $x=\cot z$, 
substituting the above factorization ansatz
into Eq.~(\ref{Schroed}), and a subsequent division by $(1+x^2)^{(1-
\beta)/2}$ one finds 
\begin{eqnarray}
&&(1+x^2)
\frac{d^{\, 2}\ C^{(\beta, \alpha) } (x)}{d\ x^2}+
2\left({\alpha\over 2}+\beta x\right) 
{d\ C^{(\beta ,\alpha ) } (x) \over d\ x }\nonumber\\
+{\Big(}(-\beta(1-\beta)-a(a+1)) &+&{(-\alpha(1-\beta)+2 b)x + \left(
\left({\alpha\over 2}\right)^2-(1-\beta )^2+
\epsilon_m\right)\over 1+x^2}{\Big)}
C^{(\beta ,\alpha )} (x)  =  0\ . \label{Sch-RMt4}
\end{eqnarray}
This equation is suited for comparison with 
\begin{equation}
(1+x^2)
\frac{d^{\, 2}\ {\mathcal R}_m^{(\beta, \alpha) }(x)}{d\ x^2}+
2\left({\alpha\over 2}+\beta x\right) 
{d\ {\mathcal R}_m^{(\beta ,\alpha ) }(x) \over d\ x }\
-m(2\beta +m-1){\mathcal R}_m^{(\beta ,\alpha )}(x)=0\, ,
\label{new_pol}
\end{equation}
which being of the form of the text-book hypergeometric equation
\cite{handbook},\cite{textbook},\cite{Dennery}  can be cast into the
self-adjoined Sturm-Liouville form given by
\begin{eqnarray}
 s (x){{d^{\, 2}{\mathcal R}^{(\beta ,\alpha)}_m(x)} \over {d\ x^2}}+
{1\over {w(x)}}\left({{d\ s(x)w(x)}\over {d\ x}}\right)
{d\ {\mathcal R}^{(\beta ,\alpha)}_m(x)
\over d\ x}+\lambda_m \ {\mathcal R}^{(\beta ,\alpha)}_m(x)&=&0\ , 
\label{d2-R2}\\
s(x)=1+x^2, \quad 
w^{(\beta,\alpha)}(x)=
s(x)^{\beta -1}e^{-\alpha\cot ^{-1}x},\,\,
\lambda_m=-m(2\beta +m-1),\,\, -\infty <x<\infty . &&
\label{weight_funct}
\end{eqnarray}
However, while the standard textbooks consider exclusively   
$s(x)$ functions which are at most  second order 
polynomials of {\it real roots\/}, in which case
\begin{equation}
 w(a)s(a)x^l= w(b)s(b)x^l=0 \ , \quad 
\forall l=\mbox{\footnotesize integer,}
\label{rule_tobreak}
\end{equation}
holds valid, the roots of $s(x)$ in Eq.~(\ref{new_pol})
are {\it imaginary}. In the former  case it is well known that
\begin{itemize}
\item ${\mathcal R}^{(\beta ,\alpha)}_m(x)$ would be polynomials of 
order $m$,  
\item $\lambda_m$ would satisfy 
\begin{eqnarray}
\lambda_m&=&-m\left(K_1{{d\ {\mathcal R}^{(\beta ,\alpha)}_1(x)} 
\over {d\ x}}+
{1\over 2}(m-1) {{d^{\, 2} s(x)}\over 
{d\ x^2}}\right)\, ,\nonumber\\ 
 \label{lamb}
\end{eqnarray} 
with $K_m$ being the ${\mathcal R}^{(\beta ,\alpha )}_m(x)$ 
normalization constant,
\item the first order polynomial would be defined as
\begin{equation}
{\mathcal R}^{(\beta ,\alpha )}_1 (x) ={1\over {K_1w(x)}}
\left({{d\ s(x)w(x)}\over {d\ x}}\right)\, ,
\label{F1}
\end{equation}
\item the latter relation would generalize to any $m$ via 
the so called Rodrigues formula 
\begin{equation} 
{\mathcal R}^{(\beta ,\alpha)}_m(x)=
\frac{1}{K_mw(x)}{d^m\over d\ x^m}(w(x)\ s(x)^m) \ ,
\label{Rodrigues-0}
\end{equation}
\item  $w(x)$ would be the weight-function with respect to
which the ${\mathcal R}^{(\beta ,\alpha)}_m(x)$ polynomials
would appear orthogonal.
\end{itemize}
Within this context the question arises whether the imaginary roots
of $s(x)=(1+x^2)$ in Eq.~(\ref{new_pol}) would prevent the 
${\mathcal R}_m^{(\beta , \alpha )}(x)$ functions from being 
real orthogonal polynomials.
The answer to this question is negative. It can be shown that
also in the latter case 
\begin{itemize}
\item the ${\mathcal R}_m^{(\beta , \alpha)}(x)$'s 
are polynomials of order $m$, 
\item they can also  be constructed systematically  from a Rodrigues formula
in terms of the respecified weight function, 

\item but only a finite number them  will be orthogonal
due to the violation of Eq.~(\ref{rule_tobreak}).
\end{itemize}
 {}From the historical perspective, 
Eq.~(\ref{weight_funct}) has first
been brought to attention 
by the celebrated English mathematician Sir 
Edward John Routh in Ref.~\cite{Routh}
(modulo the unessential difference in the argument
of the exponential from the present $\cot^{-1}$ to Routh's 
$\tan ^{-1}$), the teacher of J.\ J.\ Thomson 
and J.\ Larmor, among others famous physicists.  
Routh observed that the weight-function of the Jacobi polynomials, 
$ P^{(\mu ,\nu )}_m(x)$, takes the form of
Eq.~(\ref{weight_funct}) upon the particular complexification of the
argument and the parameters according to
$\mu \longrightarrow \eta=a+ib, \nu\longrightarrow \eta^*$, and $x\to ix$. 
From that he concluded that
$P_m^{( \eta ,\eta^\ast )}(ix)$ is a real polynomial 
(up to a global phase factor).
Later on, in 1929,  the prominent Russian  mathematician 
Vsevolod Ivanovich  Romanovski, one of the founders of the 
Tashkent University, studied few more
of their properties in \cite{Romanovski} and it was him who 
observed that only a finite number of them appear orthogonal.
While the mathematics literature is familiar with such polynomials 
\cite{NikUv}, \cite{Ismail}, \cite{Koepf}, \cite{Neretin} where they are 
referred to as finite Romanovski polynomials \cite{Zarzo}, 
or, Romanovski-Pseudo-Jacobi polynomials \cite{Lesky},
a curious omission from all the standard textbooks on mathematical physics 
\cite{handbook},\cite{textbook} takes place. 
This might be  related to circumstance that 
the physics problems which call for such
polynomials are relatively few. Recently, it has been reported in the
peer physics literature \cite{CK}, \cite{ACK} that the 
Schr\"odinger equation with the respective hyperbolic Scarf and 
trigonometric Rosen-Morse potentials is resolved exactly precisely in terms
of the Romanovski polynomials. Moreover, the latter are also relevant in   
calculation of gap probabilities in finite Cauchy random matrix 
ensembles \cite{Witte}. 
In the following, we shall adopt the notion of Routh-Romanovski polynomials
for obvious reasons.\\

\subsection{The exact spectrum}
Back to  Eq.~(\ref{Sch-RMt4}) we observe that if it is to coincide with 
Eq.~(\ref{new_pol}) then  the coefficient in front of
the $1/(x^2+1)$ term has to nullify. This restricts
the $C^{(\beta ,\alpha )}(x)$  parameters in the Schr\"odinger 
wave function to be   
\be
-\alpha(1-\beta)+2 b=0\ ,
\quad 
\left({\alpha\over 2}\right)^2-(1-\beta)^2+\epsilon=0\label{ep-b2_a2-1}\, .
\ee
With that Eq.~(\ref{Sch-RMt4}) to which one 
has reduced the original  Schr\"odinger equation  amounts to
\be (1+x^2){d^{\, 2}\ C^{(\beta ,\alpha )}(x) \over d\ x^2}+
2\left({\alpha\over 2}+\beta x\right) 
{d\ C^{(\beta ,\alpha )}(x)
\over d\ x} + (-\beta(1-\beta)-a(a+1))C^{(\beta ,\alpha )}(x) = 0\ . 
\label{Sch-RMt5}\ee

The final step is to identify the constant term in the latter equation with
the one in Eq.~(\ref{new_pol}) which amounts to a third  condition
\be 
-\beta(1-\beta)-a(a+1) = -m(2\beta+m-1)\ ,\label{b-1}
\ee
which introduces the dependence of the 
$C^{(\beta ,\alpha ) }(x)$ functions on the index $m$, i.e.
$C^{ (\beta ,\alpha )}(x)\longrightarrow C_m^{ (\beta ,\alpha ) }(x)$.
Remarkably, Eqs.~(\ref{ep-b2_a2-1}) and (\ref{b-1})
indeed allow for consistent  solutions for $\alpha$, $\beta $, and $\epsilon$
and given by (upon renaming $m$ by $(n-1))$:

\be
\beta_n=-(n+a)+1\ ,&\quad& \alpha_n={2 b\over n+a}\ ,\nonumber\\
\epsilon_n = (n+a)^2-{b^2\over (n+a)^2}\ ,
&&\quad n=m+1\, ,
\label{RMt_spectrum}
\ee
now with  $n\ge 1$.
In this way Eq.~(\ref{RMt_spectrum}) provides 
the exact tRM spectrum.
In effect, the polynomials  that define the exact solution of the
Schr\"odinger equation with the trigonometric Rose-Morse potential
turn out identical to the Routh-Romanovski polynomials
however with indices that depend on $n$. 
As we shall see below, this circumstance will become of crucial 
importance in allowing for an {\it infinite \/} number of orthogonal
polynomials (as required by the infinite depth of the potential) and thus 
for avoiding the finite orthogonality of the bare Routh-Romanovski
polynomials.

With that all the necessary ingredients for the solution of 
Eq.~(\ref{Sch-RMt5}) have been prepared. In now exploiting the
Rodrigues representation (when making the $n$ dependence explicit),
\be C^{(\beta_n,\alpha_n)}_{n}(x)\equiv
{\mathcal R}_n^{(\beta_n,\alpha_n)}(x)=
{1\over K_n\ w^{(\beta_n, \alpha_n)}(x)}
{d^{n-1}\over d\ x^{n-1}}\left(w^{(\beta_n,\alpha_n)}(x)\ 
s(x)^{n-1}\right) \, , \label{pol-nvo}\ee
allows for the systematic construction of 
the solutions of Eq.~(\ref{Sch-RMt5}).
Notice that in terms of $w^{(\beta_n ,\alpha_n)}(x)$ the wave function 
is expressed as
\be
R_n^{(a,b)}(\cot^{-1} x)=\sqrt{w^{\left(-(n+a)+1, \frac{2b}{n+a} \right)} (x)}
{\mathcal R}_n^{\left( -(n+a)+1 ,\frac{2b}{n+a} \right)}(x)\, .
\label{R_z}
\ee
\noindent
Next one can check orthogonality of the physical solutions in $z$ space
and obtain it as it should be as
\be \int_0^\pi \ R_{n}\left(z\right) 
R_{n'}\left(z\right)d z 
=\delta_{n\ n'}\, ,
\label{orth_R}\ee
The orthogonality of the wave functions $R_n(z)$  
implies  in $x$ space orthogonality of the 
${\mathcal R}_n^{(\beta_n,\alpha_n)}(x)$ polynomials with respect to
$w^{(\beta_n, \alpha_n)}(x)\frac{dz}{dx}$ due to the variable change.
As long as $\frac{d \cot^{-1} x }{dx}=-1/(1+x^2)\equiv -1/s(x)$
then the orthogonality integral for the polynomials takes the form
\be \int_{-\infty}^\infty {dx\over s(x)}
\sqrt{w^{(\beta_n,\alpha_n)}(x)}
{\mathcal R}_n^{(\beta_n,\alpha_n)}(x) 
\sqrt{w^{(\beta_{n'},\alpha_{n'})}(x)} 
{\mathcal R}_{n'}^{(\beta_{n'},\alpha_{n'})}(x)=
\delta_{n\ n'}\ .
\label{orto-1}\ee
The existence of an infinite number of orthogonal Routh-Romanovski polynomials
was made possible on cost of the $n$ dependence of the parameters which
emerged while converting the Schr\"odinger equation to the polynomial
one.

\subsection{The degeneracy in the spectra}
Inspection of Eq.~(\ref{RMt_spectrum}) reveals existence of an  
intriguing degeneracy in the tRMP spectrum. In order to 
see it let us assume that
the $a$-parameter in Eq.~(\ref{wafu_U(r)}) takes only 
integer non-negative $a=0,1,2,...$-values. In such a case, one immediately
reads off from Eq.~(\ref{RMt_spectrum}) that the energy levels
for any $\sigma =(m+1+a)$ with $\sigma =1,2,3,...$ are
$\sigma $-fold degenerate as $a$ can take all the values between
$0$ and $(\sigma -1)$ according to $a=0,1,...(\sigma -1)$ (see Fig.~3).
\begin{figure}
\begin{center}
\includegraphics[width=70mm,height=70mm]{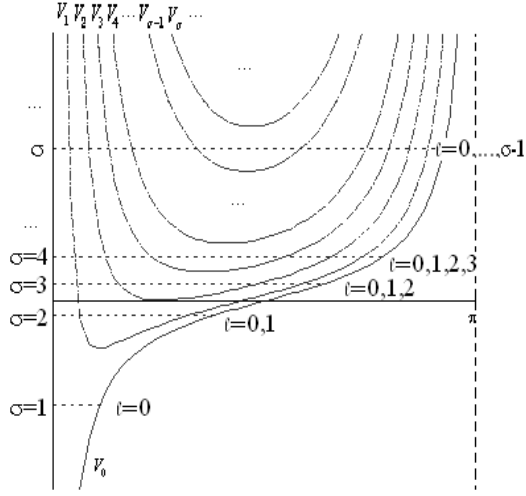}
\caption{Degeneracy of energy levels of same $\sigma $ but different
angular momenta in  $l$ dependent 
trigonometric Rosen-Morse potentials. The curves correspond to
$b=60$, a value fitted to the $N$ spectrum.}
\end{center}
\end{figure}
Comparison of the $a$-degeneracy  to the non-strange baryon spectra
in Fig.~1 and Eqs.~(\ref{brnsh_rules}) 
is suggestive of the idea to interpret the non-negative
integer $a$-values as  angular momenta and view the 
$\csc^2\left( \frac{r}{d}\right)$ term as
a {\bf non-standard} centrifugal barrier
\begin{equation}
a(a+1)\csc^2 \left( \frac{r}{d}\right) 
\longrightarrow \frac{l(l+1)}{\sin^2 \left(\frac{r}{d}\right)}\, ,
\quad a\equiv l =0,1,2,....
\label{nst_cfbr}
\end{equation} 
In terms of $\sigma $ the mass formula in Eq.~(\ref{mfla_prms})
translates to
\begin{equation}
\frac{
4
\left( 
M_{\sigma ,I}-M_{1,I} +
\frac{1}{4}
g_I
\right)
}{g_I}
\longrightarrow
\epsilon_\sigma =-\frac{b^2}{\sigma^2}+\sigma^2, \quad b^2=\frac{4f_I}{g_I}\, .
\label{fit_pot}
\end{equation}
Non-standard centrifugal barriers of various types have been frequently
exploited in the calculation of the spectral properties of collective nuclei.
Specifically, in Ref.~\cite{PTG_1} use has been made
of an angular-momentum dependent  potential originally suggested
by  Ginocchio in Refs.~\cite{PTG}. The non-standard centrifugal
barrier in this potential asymptotically approaches  
for certain parameter values the physical centrifugal barrier, $l(l+1)/r^2$ 
while for another set of parameters it  
becomes the P\"oschel-Teller potential. 
In our case, for small arguments the $\csc^ 2$ term also 
approaches the physical centrifugal term as evident from
Eq.~(\ref{Taylor_Coul}) and visualized by Figs.~4. 
Non-standard centrifugal barriers have the property to couple various 
multipole modes in nuclei, an example being given more recently in 
Ref.~\cite{Svetla}. 
From now onward we shall adopt non-negative integer values for the $a$ 
parameter and view the $\csc^2$ term as a non-standard centrifugal barrier
according to
\begin{equation}
V_l(r)=-2B\cot\left({r\over d}\right)+
{\hbar^2\over 2\mu d^2} l(l+1)\csc^2\left({r\over d}\right)\, .
\label{l_pot}
\end{equation}
In so doing we are defining a new angular momentum dependent potential,
$V_l(r)$, that possesses the dynamical $O(4)$ symmetry. 
Notice that this does not contradict
the statement of Ref.~\cite{Zeng} according to which
only pure or screened Coulomb-like potentials are $O(4)$ symmetric as
the theorem of Ref.~\cite{Zeng} refers to potentials with the standard
centrifugal barrier only.
The $b$ parameter in Eq.~(\ref{fit_pot}) now relates to the potential
parameter $B$ as
\begin{equation}
b={2 \mu d^2 B\over \hbar^2}\, .
\label{sh}
\end{equation}
Next we shall bring down the $a$ index, suppress the $b$ index
and change notations according to
\begin{equation}
R^{\left((a\equiv l), b\right)}_n\left(\frac{r}{d}\right)
\longrightarrow R_{\sigma l}\left(\frac{r}{d}\right)\, , \quad \sigma =n+l.
\label{new-R_nl}
\end{equation}
The single-particle wave functions within the angular dependent 
trigonometric Rosen-Morse potential are straightforwardly calculated
from Eq.~(\ref{R_z}). Below we list the first three levels for illustrative
purposes:
\begin{itemize}
\item \underline{ground state $\sigma=1$}:
\begin{eqnarray}
\mbox{\bf 1s:}\quad  R_{10}\left(
\frac{r}{d}\right)&=&
2\sqrt{b (b^2+1)\over (1-e^{-2 \pi b }) }
e^{-b \left( \frac{r}{d}\right)}\sin\left({
\frac{r}{d}}\right),
\label{gst}
\end{eqnarray}
\item \underline{first excited state, $\sigma =2$}:
\begin{eqnarray}
\mbox{\bf 2s:}\quad R_{20}\left(\frac{r}{d}\right)&=&
\sqrt{2 b \left(\left( \frac{b}{4}\right)^2+1\right)
\over (1-e^{-\pi b})}
e^{-b \left(\frac{r}{2 d}\right)}\sin
{\left(\frac{r}{d}\right)}
\left( b\sin\left(\frac{r}{d}\right)-2 
\cos\left(\frac{r}{d}\right)\right),\nonumber\\
\mbox{\bf 2p:}\quad
R_{21}\left(\frac{r}{d}\right)&=&2\sqrt{2\over 3}\sqrt{b  \left(
\left(\frac{b }{2}\right)^2+1\right)
\left(\left(\frac{b }{4}\right)^2+1\right)\over (1-e^{-\pi b}) } 
e^{-b  \frac{r}{ 2 d}}\sin^2
\left(\frac{r}{ d}\right),
\end{eqnarray}
\item \underline{ second excited state $\sigma =3$}:
\begin{eqnarray}
\mbox{\bf 3s:}\quad 
R_{30}
\left(
\frac{r}{d}
\right))&=&
\frac{2}{9\sqrt{3}}
\sqrt{
\frac{
\left(
b 
\left( 
\frac{b}{9}  
\right)^2+1
\right)
}
{\left(
1-e^{
-2\pi\frac{b}{3}
}
\right)
}}
e^{-b \frac{r}{3d}}
\sin\left(
\frac{r}{d}
\right)
\nonumber\\
&&\left(2\left(\left({b \over 3}\right)^2\sin^2\left({r\over d}
\right)-b\sin\left({r\over d}\right)\cos\left({r\over d}\right)
+\cos^2\left({r\over d}\right)\right)-1\right),\nonumber\\
\mbox{\bf 3p:}\quad
R_{31}\left(\frac{r}{d}\right)&=&\left({2\over 3}\right)^{\frac{3}{2}}
\sqrt{
b 
\left(
\left(
\frac{b }{3}
\right)^2
+1\right)
\left(\left(
\frac{b }{9}\right)^2+1
\right)
\over (1-e^{-2 \pi \frac{b }{3}}) } 
e^{-b  \frac{r}{ 3 d}}\sin^2\left({\frac{r}{ d}}\right)\left(b \sin
\left(r\over d\right)-6 \cos\left(r\over d\right)\right),\nonumber\\
\mbox{\bf 3d:}\quad
R_{32}\left(\frac{r}{d}\right)&=&4\sqrt{2\over 15}\sqrt{
b  \left(
\left(
\frac{b }{3}\right)^2+1\right)
\left(
\left(
\frac{b }{6}\right)^2+1
\right)
\left(\left(
\frac{b }{9}\right)^2+1
\right)\over (1-e^{-2 \pi \frac{b }{3}}) } 
e^{-b  \frac{r}{3d} }\sin^3\left({\frac{r}{ d}}\right).
\end{eqnarray}
\end{itemize}
In Figs.~5 we display as an illustrative example the 
wave functions for the first two $\sigma $ levels.
\begin{figure}
\begin{center}
{(a)}\includegraphics[width=70mm,height=70mm]{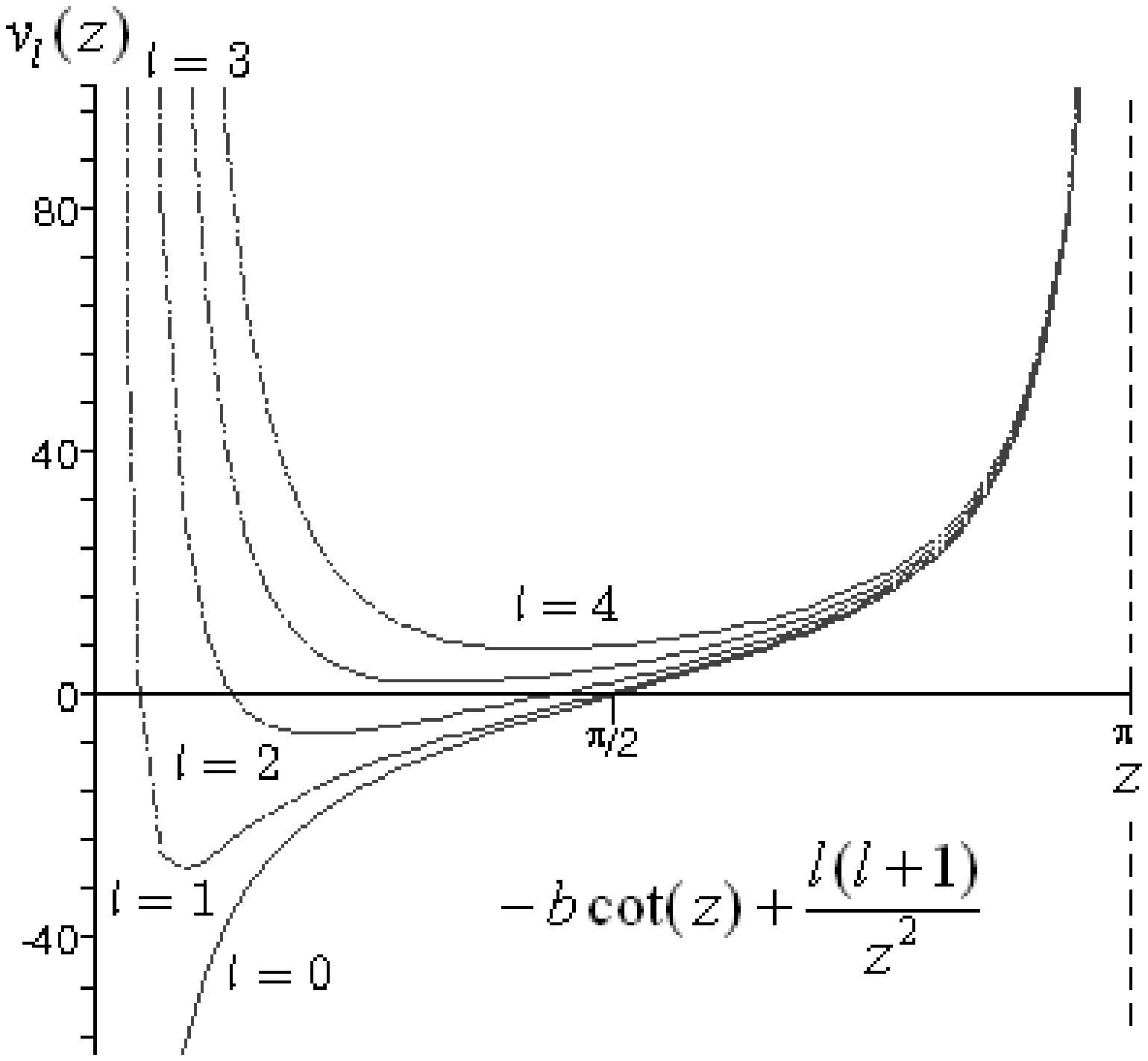}
{(b)}\includegraphics[width=70mm,height=70mm]{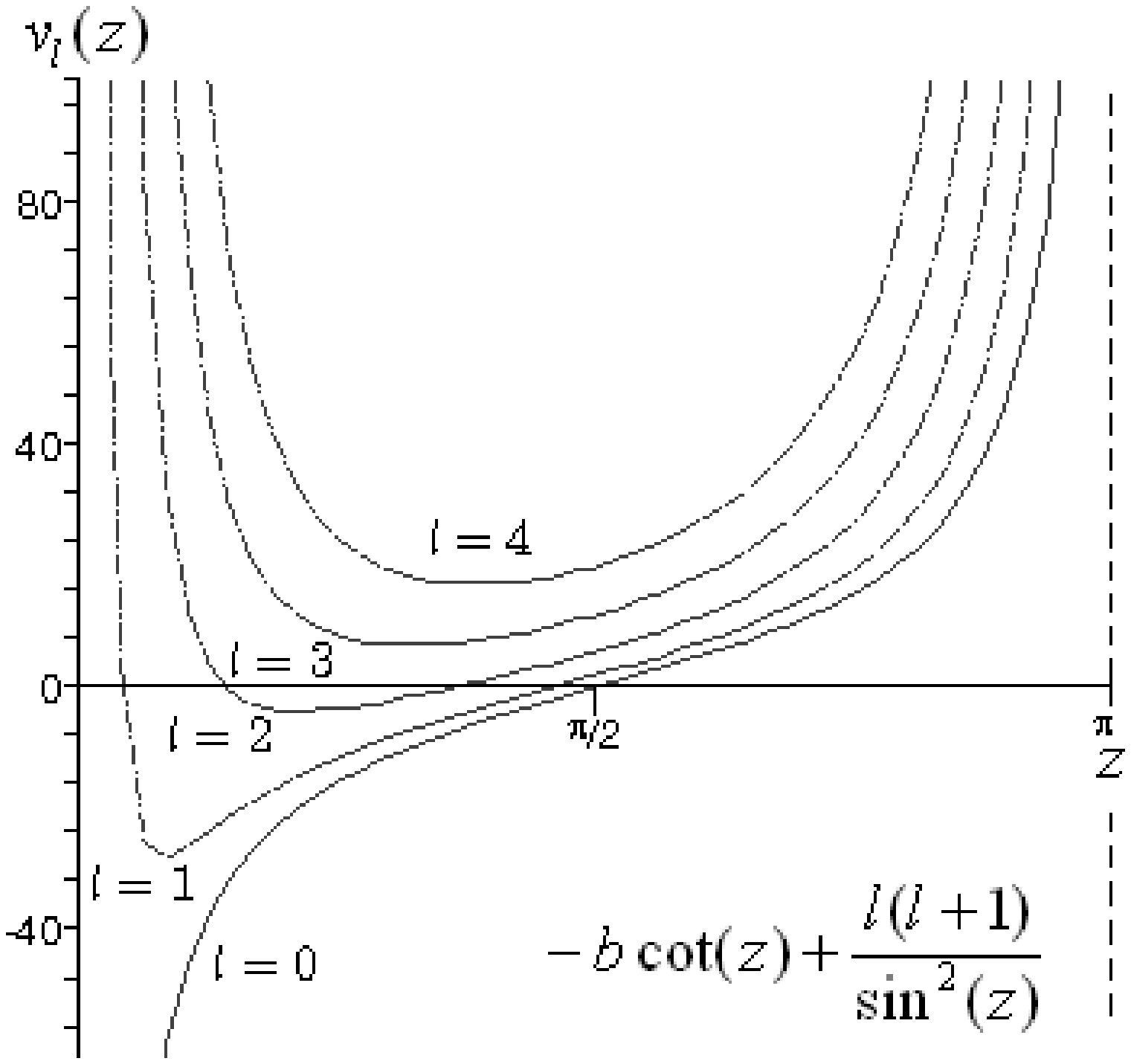}
\end{center}
\caption{The $\cot\left( \frac{ r}{d}\right) $ 
potential with the physical centrifugal barrier (left) and the 
non-standard one (right).} 
\end{figure}

\begin{figure}
\begin{center}
{(a)}\includegraphics[width=70mm,height=70mm]{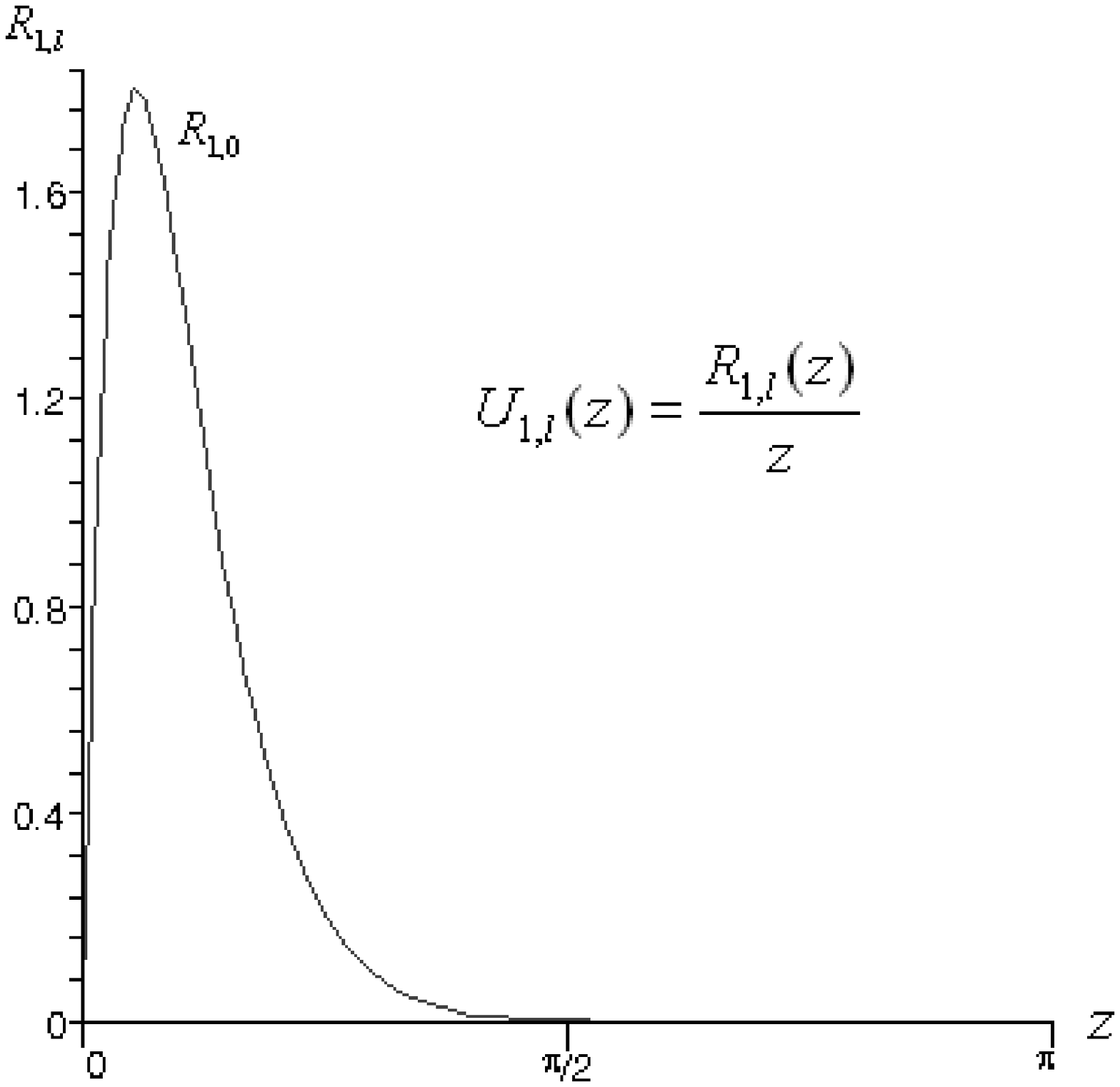}
{(b)}\includegraphics[width=70mm,height=70mm]{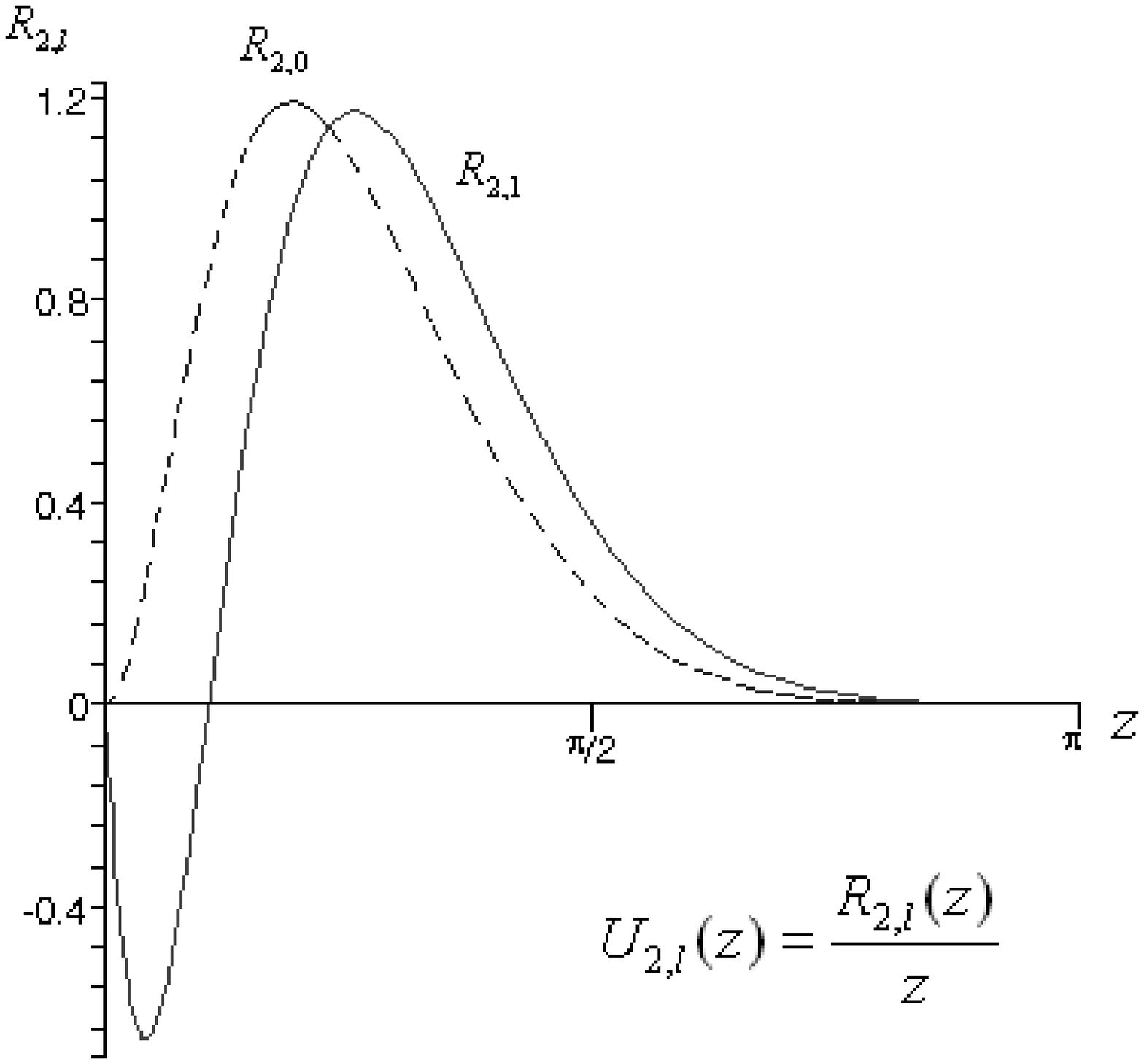}
\end{center}
\caption{ Wave functions for $\sigma =1,l=0$  (left)
and $\sigma =2, l=0,1$ (right).} 
\end{figure}

\section{Discussion and concluding remarks}
In this work we made the case that the trigonometric Rosen-Morse
potential with the $\csc^2$ term being reinterpreted as a non-standard
centrifugal barrier provides quantum numbers and
level splittings of same  dynamical $O(4)$ patterns
as observed within the $SU(2)_I\times O(4)$ classification scheme 
of baryons in the light quark sector. Due to local
$O(4)\sim O(1,3)$ isomorphism, the potential
$\left( \frac{K}{2},\frac{K}{2}\right)\otimes 
\lbrack\left(\frac{1}{2},0 \right)\oplus \left( 0,\frac{1}{2}\right) \rbrack $
levels have as a relativistic image the covariant field theoretical 
high-spin degrees of freedom, $\psi_{\mu_1...\mu_K}$. 

\noindent
We presented exact energies and wave functions of a particle within the above
potential and in so doing put it on equal algebraic footing
with the harmonic oscillator and/or the Coulomb potentials of
wide spread. 
In this fashion, 
\begin{itemize}
\item an element of covariance was brought
into the otherwise non-covariant potential picture,
\item the algebraic Hamiltonian in Eq.~(\ref{O(4)_Ham}) 
describing the $O(4)$ degeneracy 
in the $N$ and $\Delta $ spectra was translated into a potential
model of same dynamical symmetry.
\end{itemize}
The $O(4)$ degeneracy of the $N$ and $\Delta$ spectra 
seem to speak in favor of quark-diquark as leading configurations
of resonance structures. Yet, form factors are known to be more 
sensitive to configuration mixing effects and may require inclusion of
genuine three quark configurations. 
As long as the tRMP shape captures the
essential traits of the quark-gluon dynamics of QCD, we here 
consider it as a promising candidate for a realistic common quark potential
that is worth being employed in the  calculations
of spectroscopic properties of non-strange resonances.

\section*{Acknowledgments}
It is a pleasure to thank the organizers for their warm hospitality
and the excellent working conditions provided by them during the workshop.
We benefited from  extended discussions with Hans-J\"urgen Weber
and Alvaro P\'erez Raposo on various aspects  of the
Romanovski polynomials with emphasis on their orthogonality properties.

Work supported by Consejo Nacional de Ciencia y 
Tecnolog\'ia (CONACyT) Mexico under grant number C01-39280.

\end{document}